\documentclass[10pt,twocolumn,letterpaper]{article}

\usepackage[pagenumbers]{cvpr} 

\usepackage[dvipsnames]{xcolor}

\definecolor{cvprblue}{rgb}{0.21,0.49,0.74}
\usepackage[pagebackref,breaklinks,colorlinks,citecolor=cvprblue]{hyperref}
\usepackage{algorithm}%
\usepackage{algorithmicx}
\usepackage{algpseudocode}
\def\paperID{6740} 
\def\confName{CVPR}
\def\confYear{2024}
\usepackage[accsupp]{axessibility}
\title{Task-Aware Encoder Control for Deep Video Compression}
\author{
Xingtong Ge\textsuperscript{1,2}\thanks{Work was done when Xingtong Ge interned at SenseTime Research}\thanks{xingtong.ge@gmail.com} \quad \quad  
Jixiang Luo\textsuperscript{2} \quad \quad 
Xinjie Zhang\textsuperscript{3} \quad \quad 
Tongda Xu\textsuperscript{4} \quad \quad 
Guo Lu\textsuperscript{5} \\
Dailan He\textsuperscript{6} \quad \quad 
Jing Geng\textsuperscript{1}\footnotemark[2]\thanks{Corresponding author, janegeng@bit.edu.cn} \quad \quad
Yan Wang\textsuperscript{4} \quad \quad
Jun Zhang\textsuperscript{3} \quad \quad
Hongwei Qin\textsuperscript{2}\\
$^1$Beijing Institute of Technology \quad 
$^2$SenseTime Research\\
$^3$The Hong Kong University of Science and Technology \\
$^4$Institute for AI Industry Research (AIR), Tsinghua University \\
$^5$Shanghai Jiaotong University \quad 
$^6$The Chinese University of Hong Kong\\ 
}

\begin{document}
\maketitle 
\begin{abstract}

Prior research on deep video compression (DVC) for machine tasks typically necessitates training a unique codec for each specific task, mandating a dedicated decoder per task. In contrast, traditional video codecs employ a flexible encoder controller, enabling the adaptation of a single codec to different tasks through mechanisms like mode prediction. Drawing inspiration from this, we introduce an innovative encoder controller for deep video compression for machines. This controller features a mode prediction and a Group of Pictures (GoP) selection module. Our approach centralizes control at the encoding stage, allowing for adaptable encoder adjustments across different tasks, such as detection and tracking, while maintaining compatibility with a standard pre-trained DVC decoder. Empirical evidence demonstrates that our method is applicable across multiple tasks with various existing pre-trained DVCs. Moreover, extensive experiments demonstrate that our method outperforms previous DVC by about 25\% bitrate for different tasks, with only one pre-trained decoder.
\end{abstract}    
\section{Introduction}
\label{sec:intro}

\begin{figure}
  \centering
  \includegraphics[width=1.0\columnwidth]{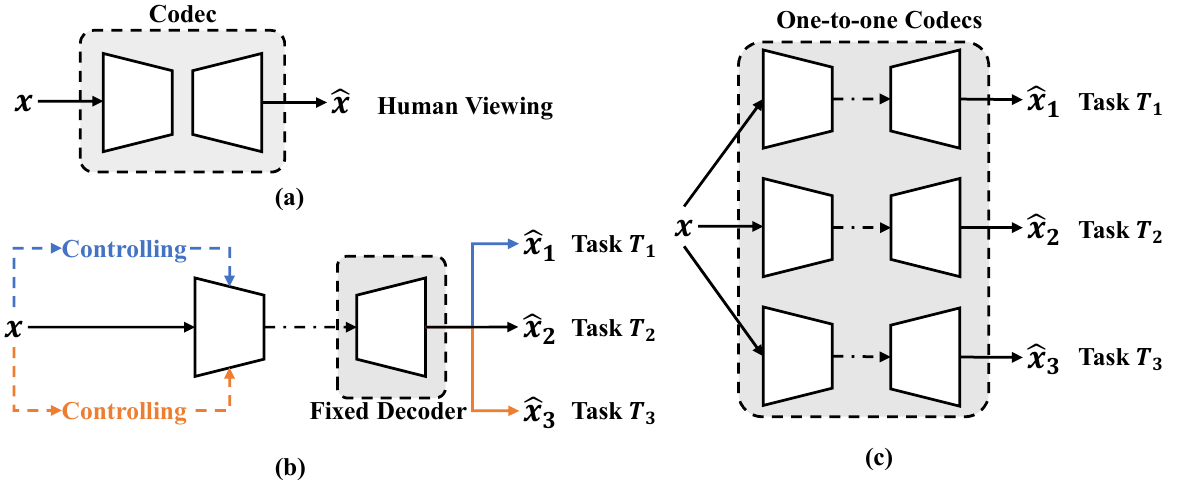}
  \caption{(a) Mainstream video codec that serves the human viewing. (b) Our controlled video codec for machine vision with fixed decoder. (c) Other video codecs for machine vison with one-to-one encoders and decoders. }
  \label{fig:fig1}
  \vspace{-0.5cm}
\end{figure}

Over the past decades, video analysis techniques have proliferated across a variety of fields, including smart cities, autonomous vehicles, and traffic surveillance. For these applications, videos are often compressed before being transmitted to cloud-based systems for further machine vision analyses, such as object detection and tracking. However, current video compression techniques, which range from traditional codecs ~\cite{h264, hevc, vvc} to recent learning-based codecs~\cite{dvc, lu2019dvc, hu2021fvc, li2021dcvc, sheng2022temporal}, primarily cater to the human visual system, as shown in Fig.~\ref{fig:fig1}(a). This specificity leads to inefficiencies, as machine vision tasks typically focus on selective semantic details and regions within the video frames, rather than the whole frames.

To tackle this problem, recent studies \cite{torfason2018towards,akbari2019dsslic,hu2020towards,li2021task,wang2021towards,yan2021sssic,fischer2021boosting,choi2022scalable,10125297,hadizadeh2023learned} have explored scalable compression for multiple tasks through different layers in image and video coding. They propose a base layer dedicated to machine vision, with an enhancement layer containing additional information for human vision. Additionally, the latest work ~\cite{tian2023non} has tried to compress semantic features at the encoder side, which serve as side information to supplement decoded images with more semantic details for machine vision at lower bitrates. Despite these advances, these methods require individually customized codecs when applied to different downstream tasks, as the ``One-to-one Codecs" shown in Fig.~\ref{fig:fig1}(c). In contrast, traditional codecs can control the encoding process for different objectives, like PSNR and SSIM optimization, using a single decoder. Inspired by this feature, 
\textbf{in this paper, we focus on how to use one pre-trained DVC-decoder to support both human and multiple machine vision tasks}.

In both residual \cite{lu2019dvc,hu2021fvc,hu2022c2f} and conditional \cite{li2021dcvc,sheng2022temporal,li2023dcvcdc} DVC series, the encoded bitstreams predominantly comprise two parts: motion and residual/contextual information. It is observed that a substantial portion of the bitrate, often exceeding 80\% in high bitrate models, is allocated to the residual/contextual information. This allocation is primarily for high-quality frame reconstruction. However, this approach is inappropriate for downstream vision tasks such as tracking and detection, which primarily concentrate on the objects and their movements rather than the full frames.

To mitigate this issue, we introduce a dynamic vision mode prediction (DVMP) module specifically tailored for machine vision tasks. This module optimizes the entropy coding process by dynamically predicting the skip/no-skip coding mode for each feature element, based on its relevance to machine vision. Specifically, it takes hyperprior data from motion or residual/contextual features within the DVC framework as the input and assesses the utility of each feature element for machine vision. By replacing non-essential elements with their predicted mean value derived from the hyperprior network, our method effectively reduces bit consumption and circumvent the entropy decoding step for these elements to expedite the decoding process. 
In this case, we generate a novel frame type that significantly reduces bitrate usage compared to traditional P frames in the Group of Pictures (GoP) structure, while maintaining the information valuable for machine vision tasks. This new frame is designated as the $P_m$ frame.

At the same time, it is important for DVC to maintain reconstruction quality because there exists significant reference relationship between frames, which is a crucial difference between image and video compression. Since our $P_m$ frame degrades the fidelity of reconstruction compared to the $P$ frame, 
we explore the reorganization of the Group of Pictures (GoP) structure. To be specific, we firstly explore to use a hand-crafted GoP structure, which already achieves significant rate-precision improvements. In the revised structure, both $P$ frames and $P_m$ frames derive their prediction from the preceding $I/P$ frame. Further, we introduce a GoP selection network, which can dynamically determine the GoP structure during the encoding process and achieves a better rate-precision trade-off.

We choose three representative residual and conditional DVC methods as touchstones in our proposed video coding for machine vision framework. It's worth mentioning that our method utilizes pre-trained DVCs without altering the weights of the original encoders and decoders. Our control happens at encoding stage, leaving the decoders' architecture intact. Moreover, when human viewing is required, we can switch back to the original encoding procedure to restore reconstruction performance. In essence, using the proposed method, we can control the encoder of a DVC to adapt for both machine and human vision requirements.

The main contributions of our work are summarised as follows: 
(1) Built upon the mainstream DVC codecs, we propose a novel video coding for machine vision framework that controls the encoder to adapt for different downstream vision tasks, such as video object detection and multi-object tracking.
(2) We employ a dynamic vision mode prediction approach to refine the original P frame, effectively reducing the bitrate while preserving critical information pertinent to vision tasks. Furthermore, we utilize a GoP selection strategy to dynamically forecast the coding GoP structure during the encoding stage, which controls the bitstream for better rate-precision trade-off in downstream vision tasks.
(3) Experiments show that our controlling method is novel and flexible for different DVC codecs, achieving up to more than 25\% bitrate savings in various downstream tasks.
\section{Related Works}

\begin{figure*}[t]
  \centering
  \includegraphics[width=1.8\columnwidth]{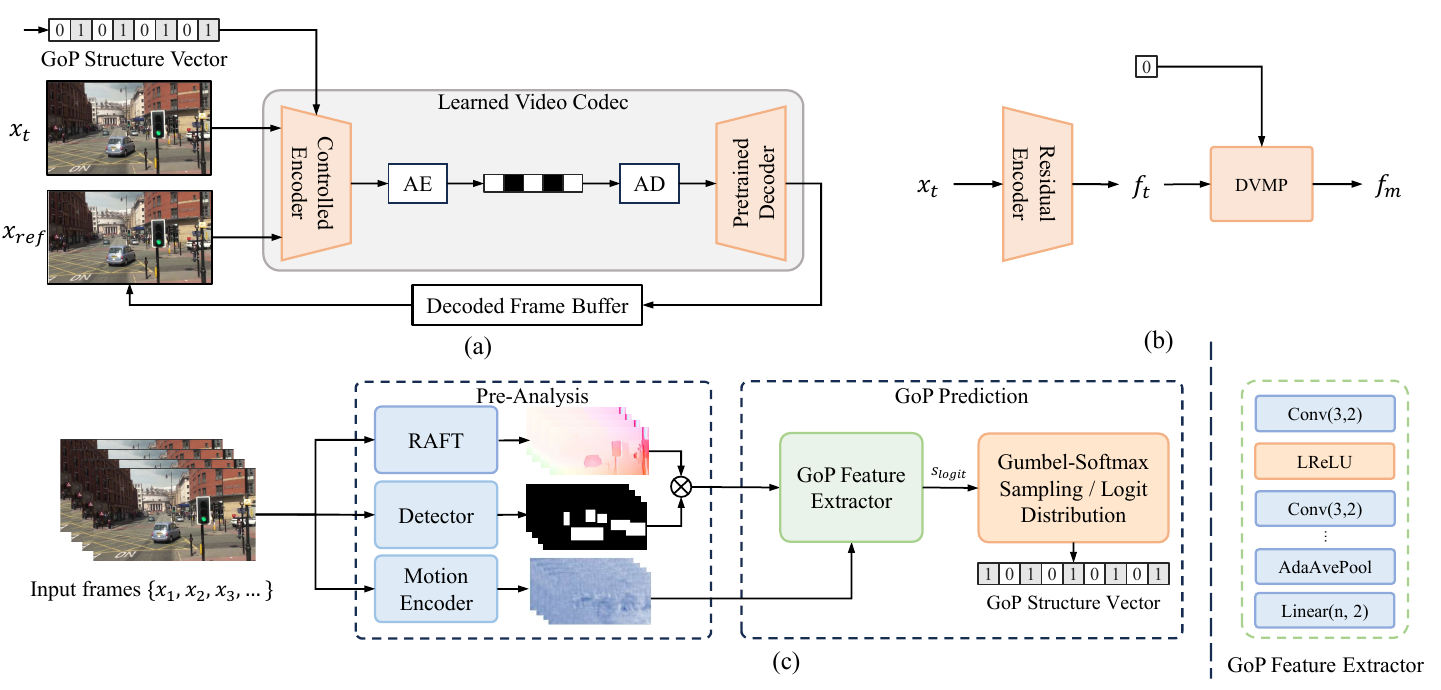}
  \caption{(a) Overview of our ``Controlling DVC for Machine" framwork. Given an input GoP, we firstly use GoP Selection network to predict the GoP sructure, then the predicted structure controls the encoding procedure to encode input frames for machine vision tasks. (b) The ``0" element controls encoder to use DVMP. (c) The GoP Selection network, including the pre-analysis stage and GoP prediction stage.}
  \vspace{-0.3cm}
  \label{fig:overview}
\end{figure*}

\subsection{Video Compression}
Previous video codecs are designed to remove spatial-temporal redundancies effectively. Traditional video coding standards, including H.264/AVC~\cite{h264}, H.265/HEVC~\cite{hevc}, and H.266/VVC~\cite{vvc}, significantly increase the compression efficiency of images and videos. Recently, a number of learning-based video codecs~\cite{dvc, lu2019dvc, hu2021fvc, li2021dcvc, sheng2022temporal,tian2021self} have been proposed, designed based on residual coding or conditional coding, achieved better and better pixel-wise signal quality metrics, e.g., PSNR and MS-SSIM~\cite{wang2003multiscale}, which mainly serve the human visual experience. Recently, there are also some generative video coding methods \cite{yang2021perceptual, mentzer2022neural} that mainly consider visual comfort and perceptual quality.

\subsection{Compression for Machine Vision} 

To facilitate the efficiency of machine vision tasks, early research works were devoted to extracting visual features from signals. Early standards, such as CDVA~\cite{cdva} and CDVS~\cite{cdvs}, suggest the pre-extraction and transportation of image keypoints to facilitate image indexing or retrieval tasks. 
With the development of deep learning, some studies began to explore connections between compression and downstream tasks, some studies~\citep{8848858, torfason2018towards,akbari2019dsslic,hu2020towards,li2021task,wang2021towards,yan2021sssic,fischer2021boosting,choi2022scalable,10125297} also focus on the joint optimization of image compression and downstream machine vision tasks by introducing a rate-distortion optimization strategy guided by downstream tasks or by adding a task-specific feature encoding stream. 
For instance, \citet{torfason2018towards} introduced a method for executing image understanding tasks, such as classification and segmentation, on compressed outputs from learning-based image compression techniques. Furthermore, Lu et al. \cite{lu2022preprocessing} enhanced traditional codecs with a preprocessing step, thereby improving codec performance for downstream vision tasks.
The field has also witnessed the emergence of self-supervised representation learning methods aimed at deriving compact semantic representations. Dubois et al.~\cite{dubois2021lossy} presented a theoretical framework suggesting that the distortion term in the lossy rate-distortion trade-off for image classification could be approximated by a contrastive learning objective. 
Feng et al.~\cite{feng2022image} proposed a method to learn a unified feature representation for AI tasks from unlabeled data. These methodologies necessitate fine-tuning the downstream models to adapt to the learned features.

Despite advancements in image coding for machine applications, reconstructing high-fidelity video from extracted features remains a formidable challenge due to the critical inter-frame reference relationships. Addressing the requirements of both machine and human vision, Tian et al.~\cite{tian2023non} introduced a self-supervised edge representation as a semantic intermediary, which preserves the video's semantic structure. Furthermore, Lin et al.~\cite{lin2023deepsvc} developed a scalable video coding framework tailored for machines, segregating semantic features for machine analysis and human viewing into separate bitstreams.

Most existing approaches~\cite{tian2024coding,lin2023deepsvc,tian2023non}necessitate the use of distinct encoders and decoders for various tasks, which complicates and burdens the coding pipelines. Our methodology advances the ``Coding for Machine" paradigm by focusing on the encoder side. We present a versatile framework capable of accommodating both human and machine vision demands without the need to modify the pre-trained decoders of existing video codecs.

\subsection{Compressed Video Analysis/Understanding}

There are also amounts of works performing video analysis tasks~\cite{liu2022real, wang2019fast, li2022end}, such as image recognition, action recognition~\cite{wu2018compressed, shou2019dmc} and multiple object tracking (MOT)~\cite{khatoonabadi2012video, dai2022survey}, in the compressed video domain. However, these methods focus on developing video analysis models that better leverage the partially decoded video stream, such as the motion vector. In contrast, our work focuses on the coding procedure, specifically the encoding procedure.

\section{Method}

\subsection{Overview}

Our coding framework is illustrated in Fig.~\ref{fig:overview} (a). Let $\mathcal{X}=\{\boldsymbol{x}_1, \cdots, \boldsymbol{x}_T\}$ represent the video sequence. The framework categorizes the $P$ frames into two types: original $P$ frames and machine vision-specific $P_m$ frames. The categorization of frame $\boldsymbol{x}_t$ is determined by the GoP Selection Network, where a value of $1$ signifies a $P$ frame and $0$ a $P_m$ frame. The $P_m$ frames, designed to reduce the bitrate while preserving critical information for future vision tasks, are generated from the preceding $P$ frames using Dynamic Vision Mode Prediction (DVMP). To maintain the quality of the decoded video sequence, each $P$ frame is computed from the previous $P$ frame with superior visual quality, rather than from a $P_m$ frame. In scenarios requiring coding for downstream vision tasks, a machine-centric control applies the GoP selection and DVMP, altering the encoding structure to "$I, P_m, P, \cdots$". Conversely, for human viewing, a human-centric control restores the encoding structure to its original form, "$I,P,P,P, \cdots$".

\subsection{Dynamic Vision Mode Prediction}

In our coding framework, we aim to minimize the video sequence bitrate. Initial analyses of residual video codecs reveal that they allocate a substantial portion of the bitrate—exceeding 80\% in high bitrate configurations—to encode and transmit residual information. This approach ensures high-quality visuals for human perception, as measured by metrics such as PSNR or MS-SSIM~\cite{wang2003multiscale}. However, this method introduces significant redundancy, especially since downstream tasks like video object detection or tracking predominantly concentrate on specific regions of interest rather than the entirety of the frame. Consequently, it is crucial for our study to develop an approach that reduces this redundancy in the coding bitstream.

\begin{figure}
  \centering
  \includegraphics[width=0.9\columnwidth]{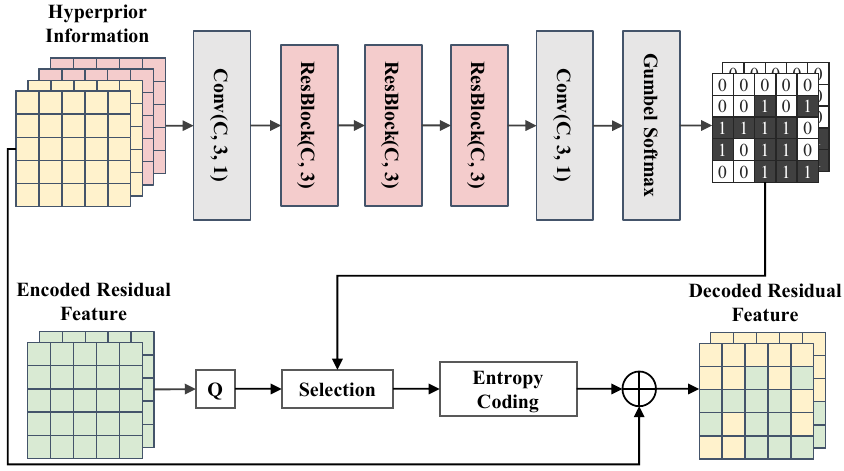}
  \caption{Hyper-prior guided Dynamic Vision Mode Prediction network.}
  \label{fig:modeprediction}
\end{figure}

Inspired by Hu et al.~\cite{hu2023complexity}, who developed a mode prediction technique specifically designed for the human visual system within a slimmable encoder and decoder, we introduce the Dynamic Vision Mode Prediction (DVMP) module. This innovative module autonomously selects the optimal coding mode and decides whether each feature element should be encoded and transmitted, as illustrated in Fig.~\ref{fig:modeprediction}.

To elucidate the functionality of our proposed Dynamic Vision Mode Prediction (DVMP) module, we utilize the encoded residual feature $m_{t}$ as an illustrative example. Assume $m_{t}$ possesses dimensions $c \times h \times w$
, indicating $c$ channels each with a spatial dimension of 
$h\times w$. The hyperprior network then predicts the mean and variance for each element, resulting in dimensions of $2c \times h \times w$. The mode prediction network's architecture, detailed in the upper branch, consists of two convolution layers and three ResBlocks. To render the ``mode prediction" process differentiable, we employ the Gumbel Softmax technique during training to ascertain the skip mode for each encoded residual feature element. This module generates a binary mask for the feature map $m_t$, where a ``1" denotes elements imperative for machine vision that require entropy coding, and a "0" signifies elements either irrelevant for machine vision or accurately predictable by the hyperprior network, thus streamlining the entropy coding process and bitrate reduction. 
Note that the showed architecture is not suitble for prior models with autoregressive components, like which in DCVC~\cite{li2021dcvc}. However, we can extend the module into an autoregressive way to enable the support for DCVC. The detailed structure is shown in the supplementary material. 

To optimize the DVMP module, the loss function can be given by:
\begin{equation}
  \begin{aligned}
    \mathcal{L}_{m} &= R + \lambda_{m} \mathcal{L}_{t}
    \label{eq:loss_mode_prediction}
  \end{aligned}
\end{equation}
where $R$ denotes the coding bitrate and $\mathcal{L}_{t}$ denotes the loss of downstream vision task, respectively. 
$\lambda_{m}$ is a hyper-parameter used to control the trade-off.

\begin{figure}
  \centering
  \includegraphics[width=1.0\columnwidth]{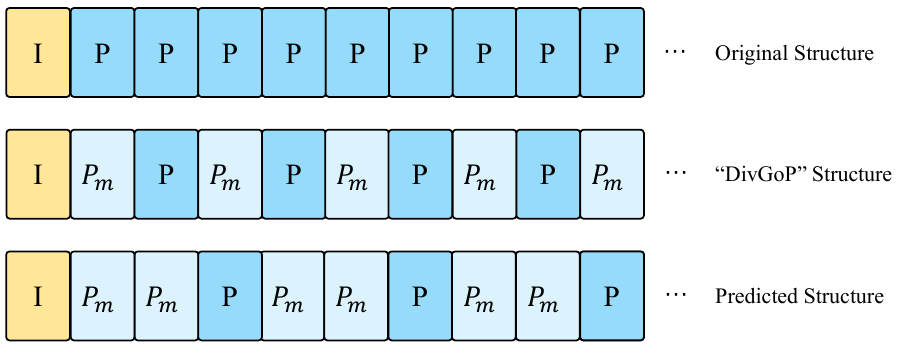}
  \caption{Different coding GoP structures. The original structure consists of I and P frames. In the middle, the hand-crafted structure for machine vision consists of I, P and $P_m$ frames which are arranged alternately. In the predicted (GoP selected) structure, the type of frames are predicted by the GoP selection network, targeting on better bitrate and machine vision performance trade-off.}
  \label{fig:gopstructure}
  \vspace{-0.3cm}
\end{figure}

\subsection{DivGoP → GoP Selection}

\begin{figure}[t]
  \subfloat
  {\includegraphics[scale=0.22]{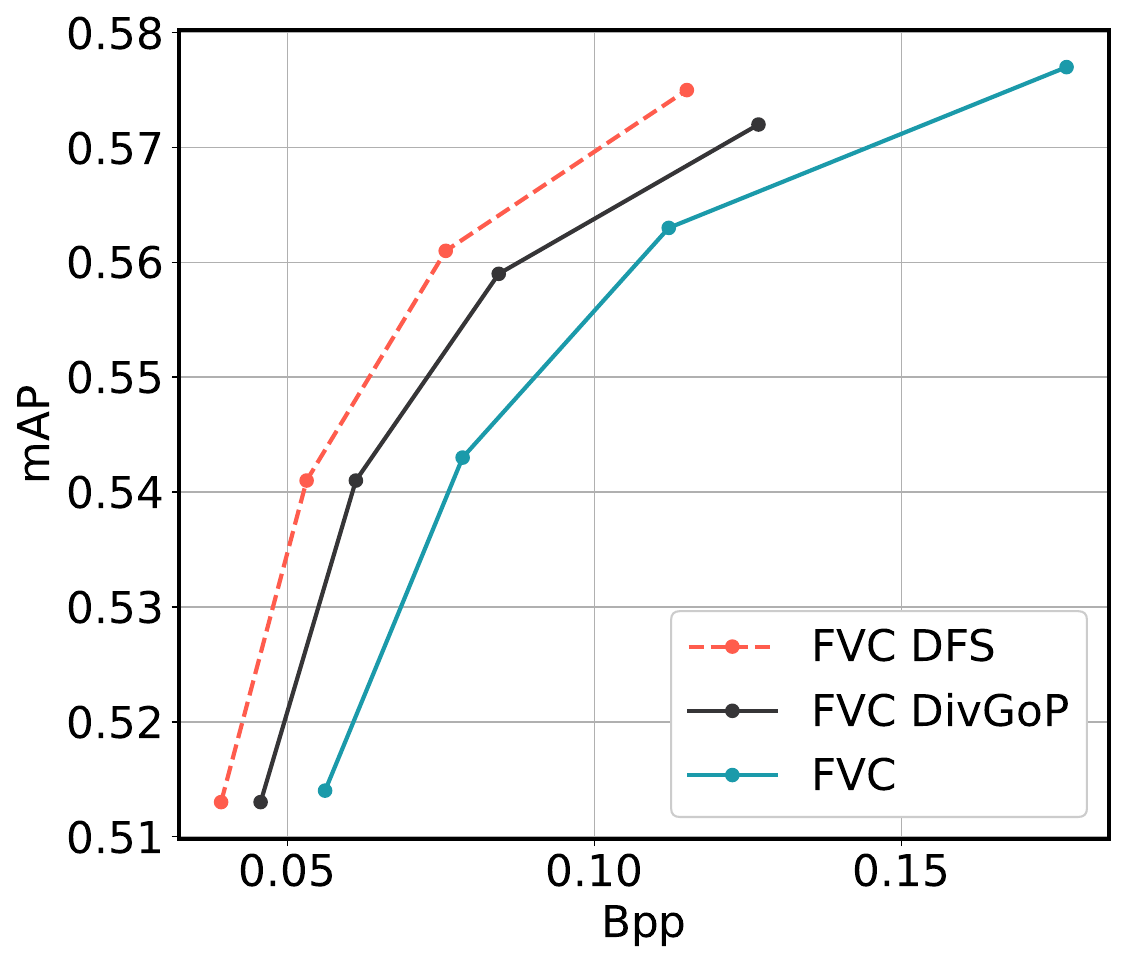}}
  \subfloat
  {\includegraphics[scale=0.22]{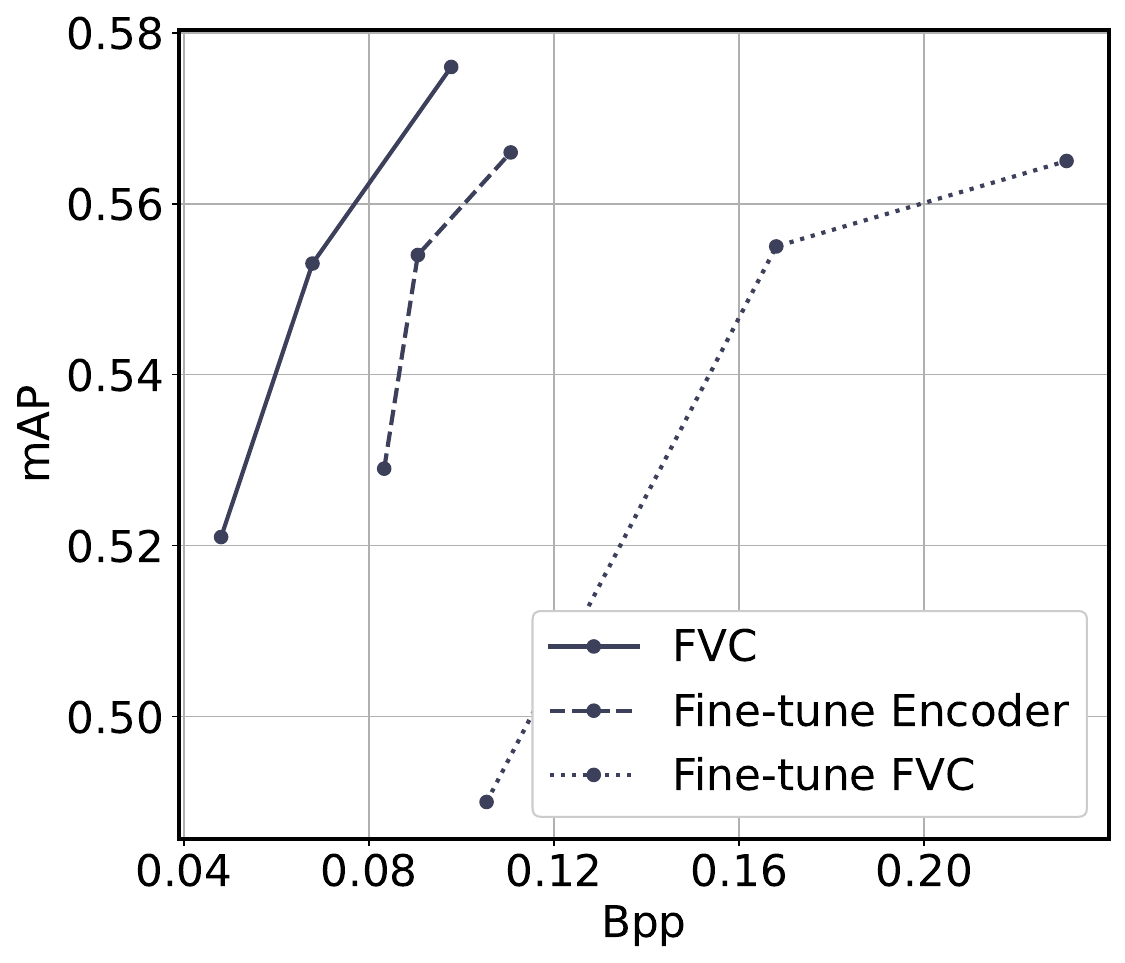}}
  \caption{Left: Using DFS to search for the near-optimal GoP structure for Bpp-mAP. Right: Results of simply fine-tuning FVC }
  \label{fig:dfs}
  \vspace{-0.1cm}
\end{figure}

Obtaining the $P_m$ frames, we firstly use a hand-crafted structure in a GoP, where the $P_m$ frames and $P$ frames are arranged alternately (referred as ``DivGoP" and shown in Fig.~\ref{fig:gopstructure}). It is observed that this arrangement can bring a certan degree of bitrate saving in detection, since the $P_m$ frames cost much less bitrate than $P$ frames while maintaining the motion and machine vision information. 
Experimental findings, as depicted in the ``FVC DivGoP" curve within Fig.~\ref{fig:dfs}, demonstrate that the "$I, P_m, P, \cdots$" structure markedly achieve about 18.6\% bitrate saving.

The ``DivGoP" configuration, while methodically structured, may not be universally optimal due to the variability in motion and object information across different videos, which could necessitate distinct GoP structures. For instance, videos characterized by rapid target movement might demand a higher frequency of $P$ frames to preserve the quality of reconstruction, as an excess of $P_m$ frames may lead to degradation that negatively affects subsequent frames. Conversely, in scenarios where the camera remains static or the target movement is minimal, the introduction of more $P_m$ frames could be feasible without significantly impacting the reconstruction quality of following frames. Stimulated by this notion, our research endeavors to tailor the GoP structure dynamically, enhancing codec performance for machine vision tasks. To approach this issue, we adopt FVC~\cite{hu2021fvc} as the foundational codec, select a GoP size of 10, and address the detection task utilizing the MOT Dataset. The optimization problem is thus defined:
\begin{equation}
\begin{aligned}
\label{equation:gop_selection_problem}
    \underset{\theta}{\text{arg min}} \quad R(\theta) + \lambda_p L_{\text{det}}(\theta)
\end{aligned}
\end{equation}

where $R$ denotes the bitrate, $L_{det}$ denotes the detection loss, $\lambda_{p}$ is a weighting factor that balances the effects of bitrate and detection loss, and $\theta$ denotes the setting of GoP structure here.

Obtaining a GoP, each of the 9 predicted frames can be either $P$ frame or $P_m$ frame, which can be dynamically decided. To explore the optimal GoP structure for target function \ref{equation:gop_selection_problem}, and evaluate the performance gap between DivGoP and optimal GoP structure, we use a deep-first-searching (DFS) algorithm, where we use different $\lambda$ for different bitrate models. The DFS algorithm searches for the structure that makes the target function smallest for each GoP. Results are shown in the ``FVC DFS" curve of Fig.~\ref{fig:dfs}. The application of the DFS algorithm led to a substantial increase in Bpp-mAP performance,, achieving about 32.3\% bitrate saving in terms of the BD-RATE metric, while the ``DivGoP" method achieves about 18.6\% bitrate saving. Similar conclusions can also be observed on DCVC~\cite{li2021dcvc} and TCM~\cite{sheng2022temporal}, reinforcing our hypothesis that distinct videos necessitate unique GoP structures tailored for machine vision tasks.
However, the DFS algorithm is extreamly time consuming, and the labels do not exist during real inference process, making the DFS not avaliable to be applied in real time encoding. How to further find an ``adaptive GoP structure" in an available time complexity is worth exploring.

From this perspective, we introduce our GoP selection method to adaptively predict the GoP structure at encoding stage, which has a low time complexity and is available in application.
Illustrated in Fig.~\ref{fig:overview} (c), our GoP selection network operates in two phases: the pre-analysis and the GoP prediction stages. Initially, input frames undergo pre-analysis, where a detector identifies objects and generates an overlapping mask for each frame, while the RAFT~\cite{teed2020raft} optical flow estimation method calculates optical flow features for each predicted frame using its adjacent predecessor. These features are then refined by masking with the object bounding box mask to produce masked optical flow features, alongside motion feature priors (mean and variance, averaged by channel) derived from the codec's motion encoder. Subsequently, these processed features are input into the GoP prediction network, which employs a feature extractor with convolution blocks for downsampling and aggregation, followed by adaptive average pooling and linear layers that output a normalized $s_{logit}$ for this GoP, which comprises two elements that sum to one. The averaged detection confidence information from the detector is also injected into the linear layers as meta data.

Upon acquiring the $s_{\text{logit}}$, we employ the Gumbel softmax sampling technique throughout the training phase to iteratively generate a binary value for each frame. This process culminates in the formation of a GoP structure vector, where the binary value---$0$ for a $P_m$ frame and $1$ for a $P$ frame---specifies the type of each frame within the GoP. During inference stage, we apply a softmax operation on $s_{\text{logit}}$ to derive probabilities, which are then used to evenly distribute $1$s and $0$s, thereby determining the actual encoded GoP structure. Similar to our mode prediction network, the loss function for optimizing GoP structure selection network is given by:

\begin{equation}
  \begin{aligned}
    \mathcal{L}_{g} &= \Bar{R} + \lambda_{g} \Bar{\mathcal{L}_{t}}
    \label{eq:loss_gop_selection}
  \end{aligned}
\end{equation}
where the $\Bar{R}$ and $\Bar{\mathcal{L}_{t}}$ denote the GoP-wise average coding bitrate and average loss of downstream vision task.

\section{Experiments}

\begin{figure*}[t]
  \subfloat[]
  {\includegraphics[scale=0.31]{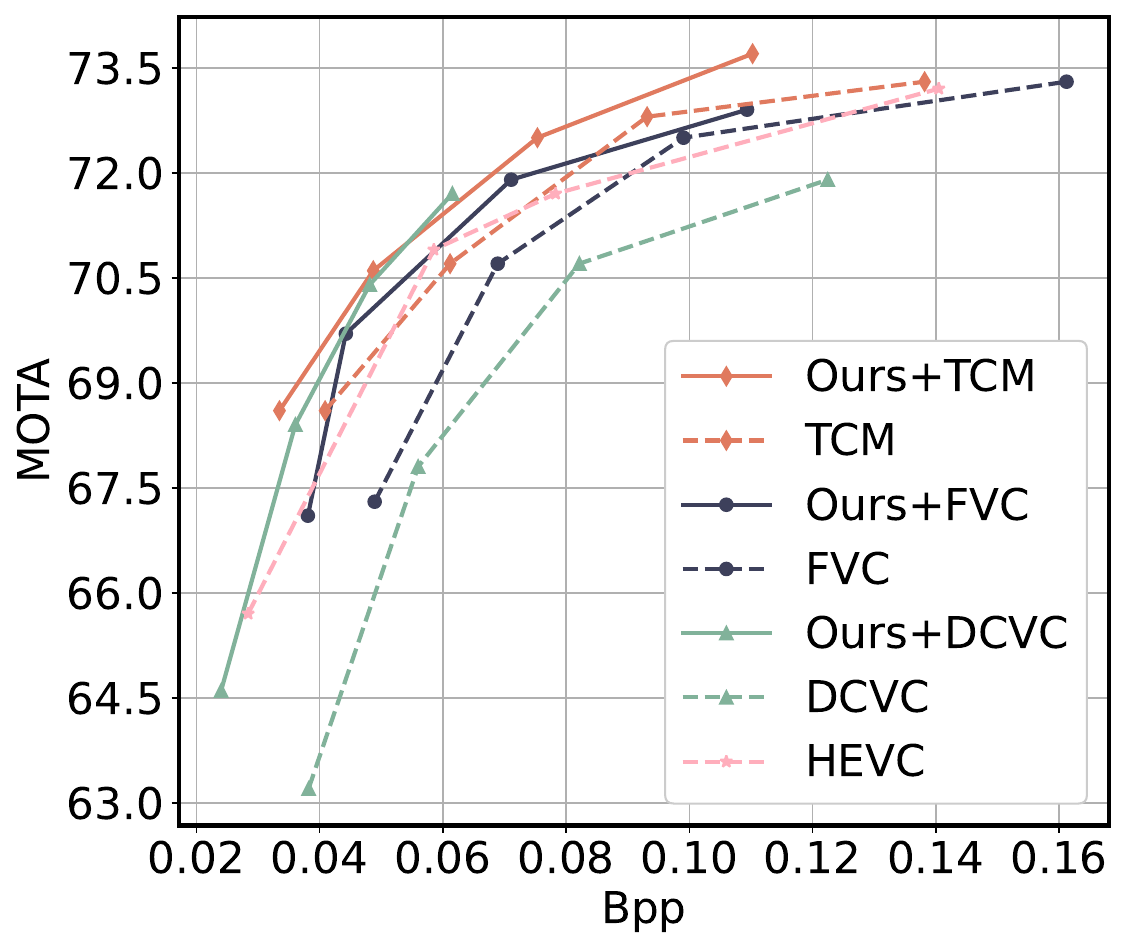}}
  \subfloat[]
  {\includegraphics[scale=0.31]{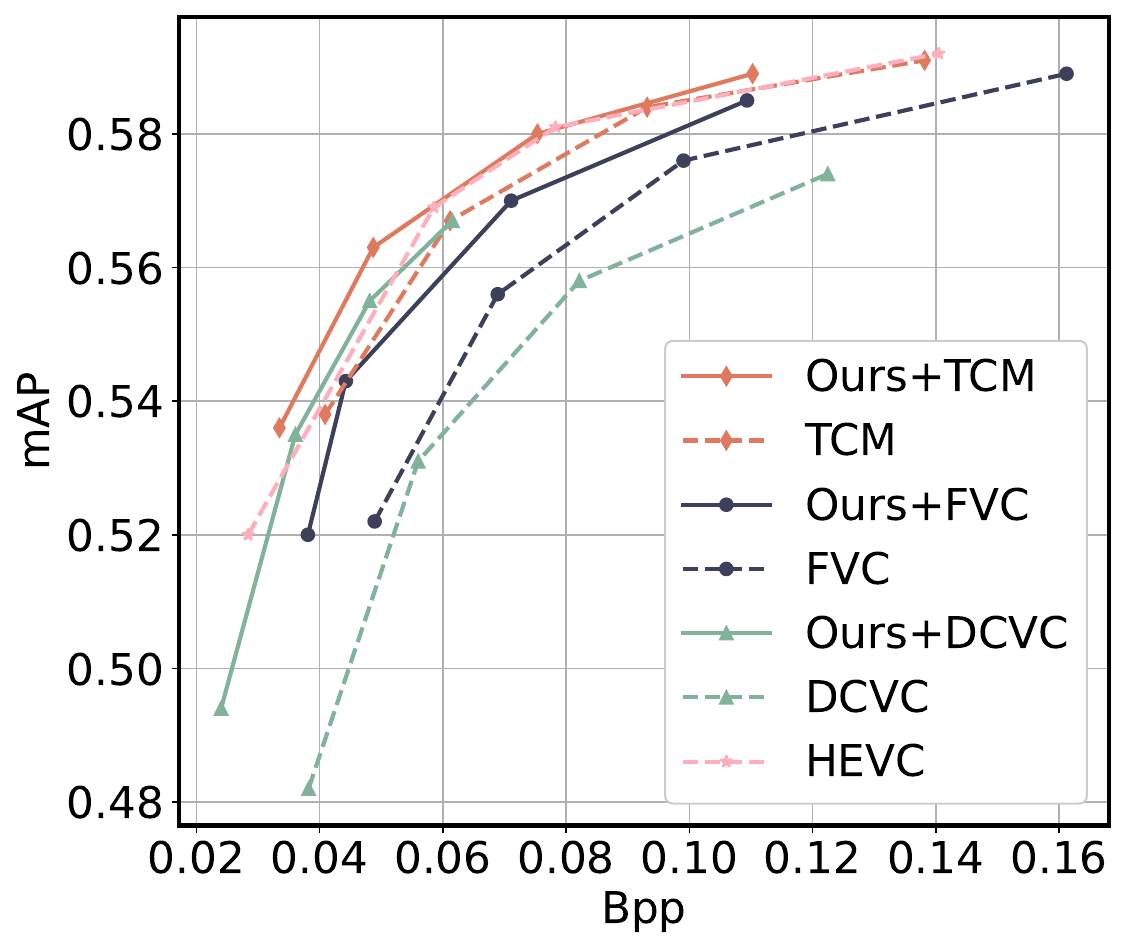}}
  \subfloat[]
  {\includegraphics[scale=0.31]{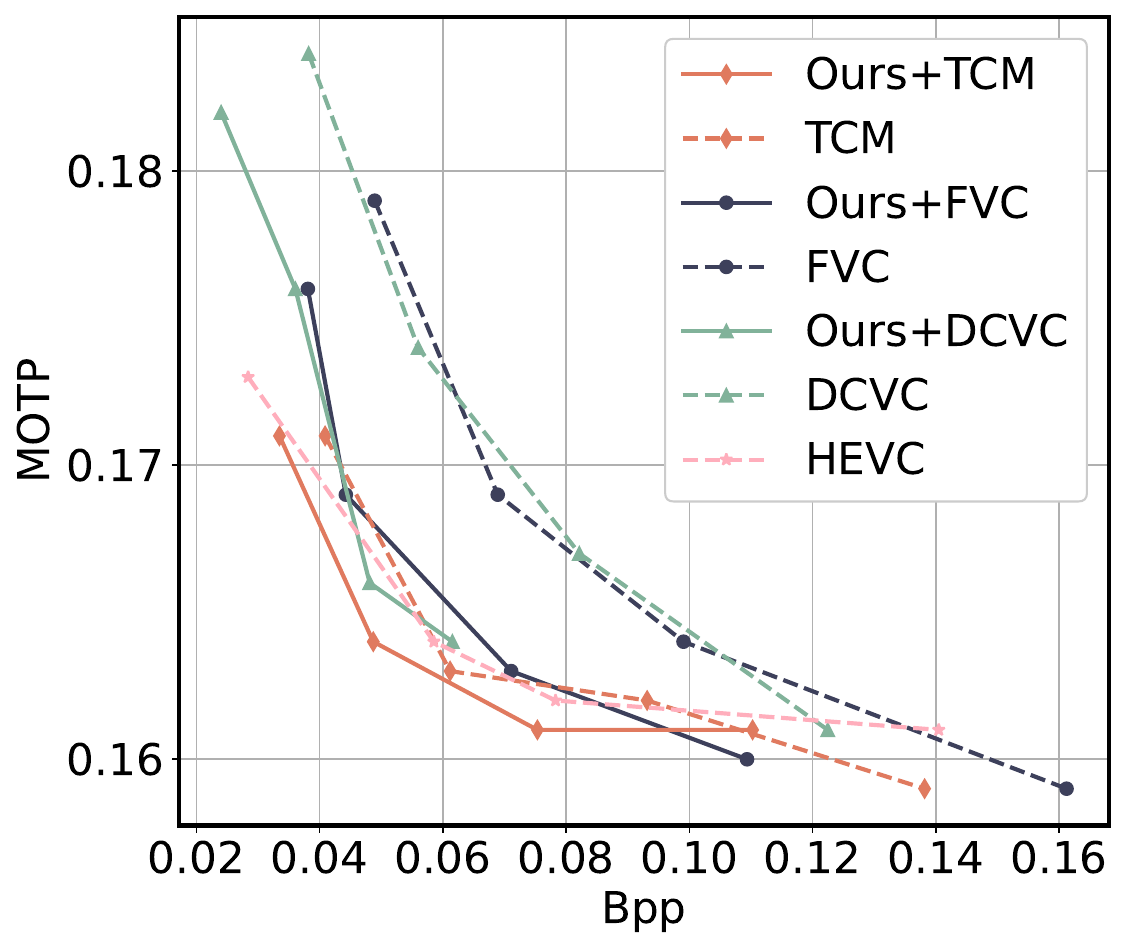}}
  \\
  \subfloat[]
  {\includegraphics[scale=0.31]{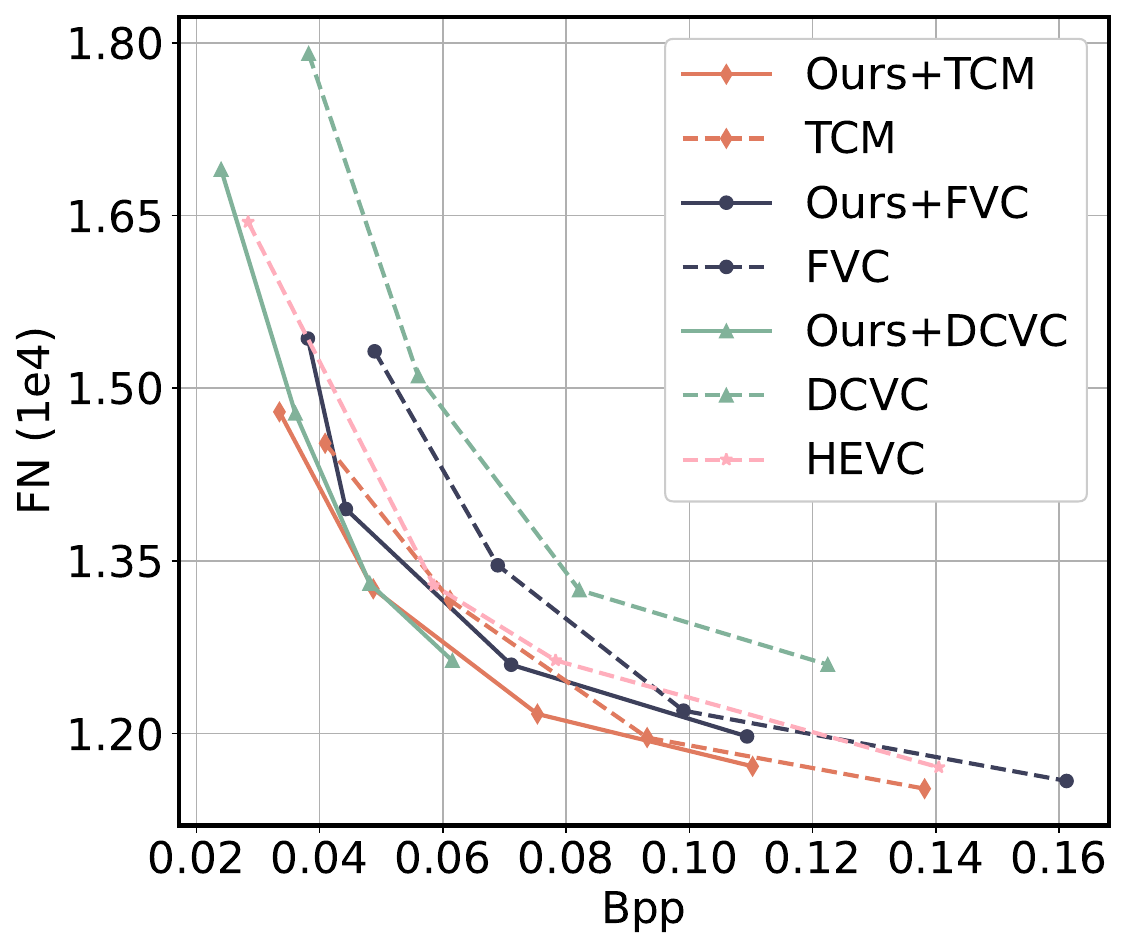}}
  \subfloat[]
  {\includegraphics[scale=0.31]{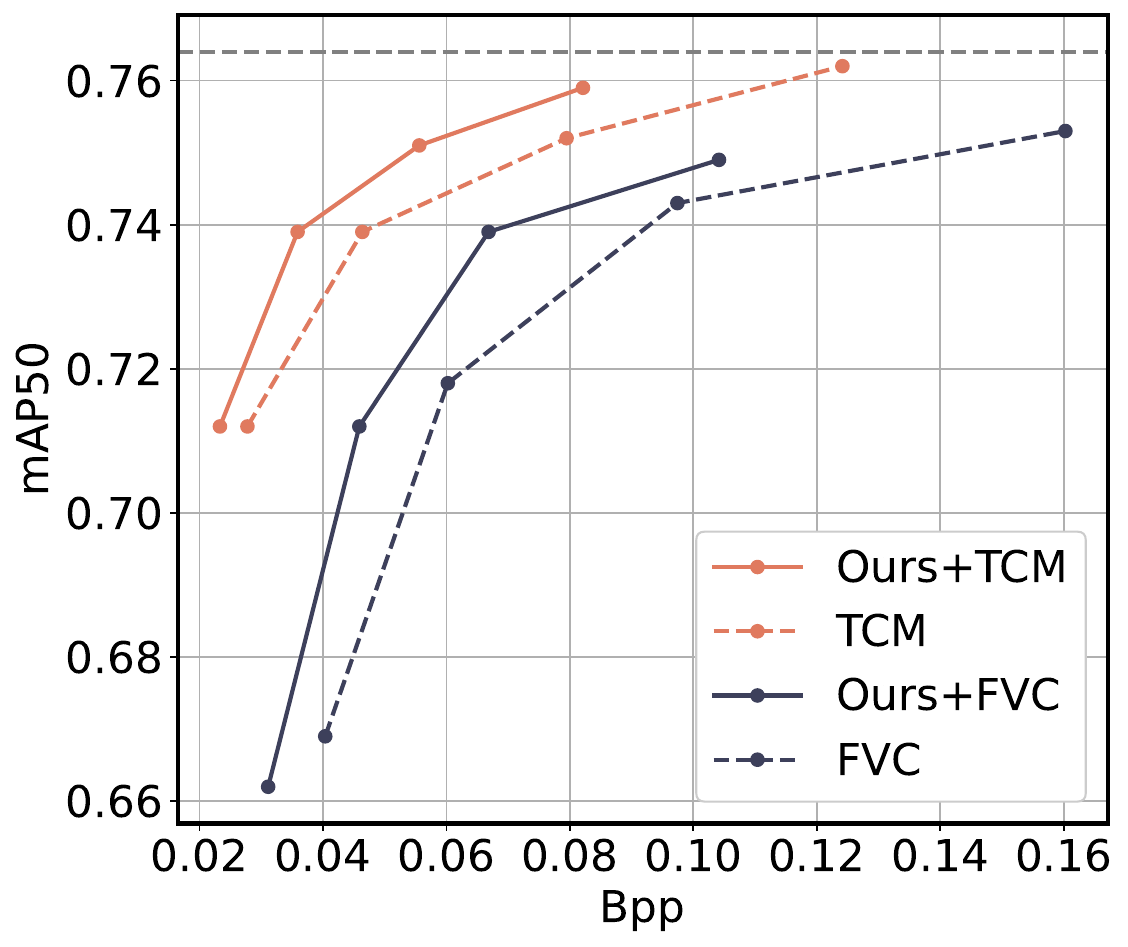}}
  \subfloat[]
  {\includegraphics[scale=0.31]{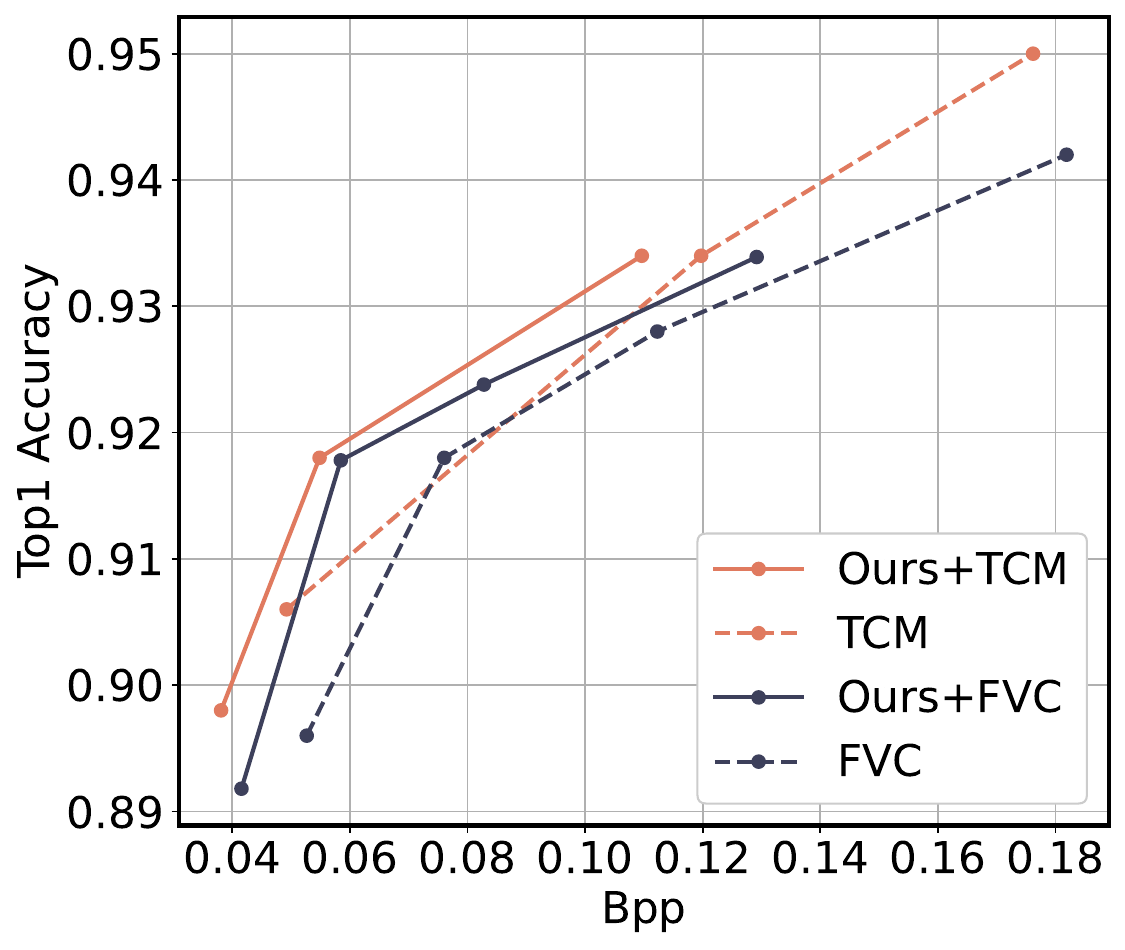}}
  \caption{(a) Bpp-MOTA curves on MOT Dataset. (b) Bpp-mAP curves on MOT Dataset. (c) Bpp-MOTP curves on MOT Dataset. (d) Bpp-FN curves on MOT Dataset. (e) Bpp-mAP50 curves on ImageNet VID Dataset using YOLOV~\cite{shi2023yolov}. (f) Bpp-Top1 Accuracy curves on UCF101 Dataset.}
  \label{fig:mot}
  \vspace{-0.2cm}
\end{figure*}

In this section, we describe the dataset settings, downstream task models, and evaluation metrics. We then provide empirical evidence underscoring the effectiveness of our proposed method across multiple tasks and codecs. Lastly, we validate the contributions of the mode prediction and GoP selection modules to performance improvements through ablation studies.

\noindent\textbf{Datasets}.
For multi-object tracking (MOT) task, we choose MOT17 dataset. For video object detection (VOD) task, we adopt the Imagenet VID dataset~\cite{russakovsky2015imagenet} for training and evaluation. For video action recognition (VAR) task, we adopt UCF101\cite{soomro2012ucf101} dataset. For the data processing procedure, we downsample the short side of the evaluation videos to $512$ for MOT17 dataset and keep the original frame ratios for both training and evaluation. For the evaluation of VOD, we follow the most VOD works to resize the frames to $512 \times 512$ for training and $576\times576$ for evaluation.

\noindent\textbf{Downstream Task Models}.
For MOT task, we adpot the popular ByteTrack~\cite{zhang2022bytetrack} method as the backbone model. The model weights are provided by their official framework. For VOD task, we adpot the detector YOLOV~\cite{shi2023yolov}, and the model weights are also provided by its official framework. For the VAR task, we adpot the TSM~\cite{lin2019tsm} model as the recognition network, the weights are provided by mmaction2~\cite{2020mmaction2}. During training and evaluation of our proposed framework, the weights of the downstream task models are fixed, without any fine-tuning. 

\noindent\textbf{Evaluation Metrics}.
For compression, We use bpp (bits per pixel) to measure the average number of bits for one pixel. For the MOT task, we adopt the MOTA (multiple object tracking accuracy), MOTP (multiple object tracking precision), FN (false negative detection number) and mAP (mean average precision) metrics to evaluate the tracking and detection performance. For the VOD task, we adopt the mAP50 metric to evaluate the detection performance, following most VOD methods. For the VAR task, we adopt the Top1 Accuracy as the performance indicator. 

\noindent\textbf{Implementation Details}.
Our whole framework is implemented on PyTorch with CUDA support and trained on eight V100 or A100 GPUs. We use the Adam optimizer by setting the learning rate as 0.0001, $\beta_{1}$ as 0.9 and $\beta_{2}$ as 0.999, respectively.
The whole training process has the follow stages: Firstly, we use the loss function \ref{eq:loss_mode_prediction} to train the mode prediction network and obtain the $P_m$ frames. This stage we use single frame training. Note that for TCM, DVMP is applied to both motion and contextual information. For DCVC and FVC, DVMP is exclusively utilized for motion information, and all residual/contextual information is omitted. After that, we introduce the GoP selection network in a GOP length unit. Using the GoP size of 32 on MOT and VAR tasks, we devide the GoP into 2 mini GoPs with size of 16, and the predicted vector of each mini GoP has a length of 15. For VOD task, we directly set the GoP size as 16. Using multi-frame training and loss function \ref{eq:loss_gop_selection}, we fix the weights of mode prediction network and train GoP selection network for about 100 epochs on MOT half-train dataset and 20 epochs on ImageNet VID training dataset, respectively for multi-object tracking and video object detection tasks. The whole training procedure takes about 2 days for each task.

\subsection{Multi-Object Tracking}
In Fig.~\ref{fig:mot}, we present a comparison between our method and various codecs, specifically DCVC, FVC, and TCM. It is evident that our approach yields a markedly improved trade-off in terms of Bpp-MOTA, Bpp-mAP, Bpp-MOTP and Bpp-FN metrics. When compared to the baseline codecs FVC, DCVC and TCM, our control method results in bitrate savings of approximately 27.54\%, 41.82\% and 12.64\%, respectively, for equivalent MOTA metrics. With DCVC established as the anchor baseline, Table \ref{table:mot_bdbr} displays the BD-RATE values, where our controlled TCM, controlled DCVC and controlled FVC codecs demonstrate the best, second-best and third-best performance, respectively. Additionally, we contrast our method with the task-oriented ``One-to-one Codecs" of VCM, with detailed results provided in the supplementary materials.

\subsection{Video Object Detection}
For the evaluation of the video object detection task, we juxtapose our method with established codecs such as FVC, and TCM, with the comparative results illustrated in Fig.~\ref{fig:mot}(e). Our method is noted to achieve a superior trade-off, particularly in the Bpp-mAP50 metric, when applied to the downstream model YOLOV. For instance, in comparison to the FVC and TCM baseline codecs, our method facilitates bitrate reductions of approximately 21.76\% and 20.42\%, respectively, while maintaining the same mAP50 metric. The uncompressed video represents the upper bound for the mAP50 metric, recorded at 76.4, and is delineated by the grey line within the figure.

\begin{figure}[t]
  \subfloat
  {\includegraphics[scale=0.22]{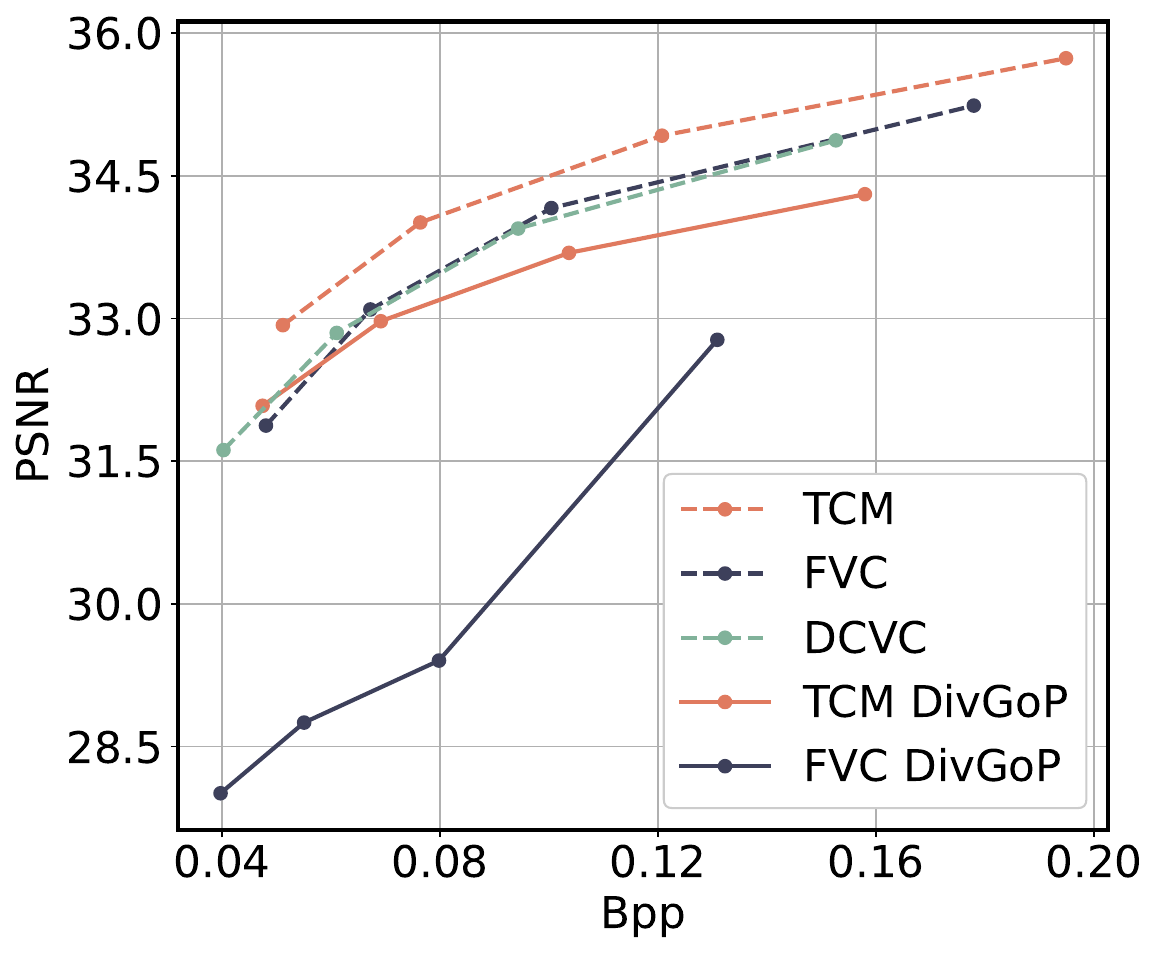}}
  \subfloat
  {\includegraphics[scale=0.22]{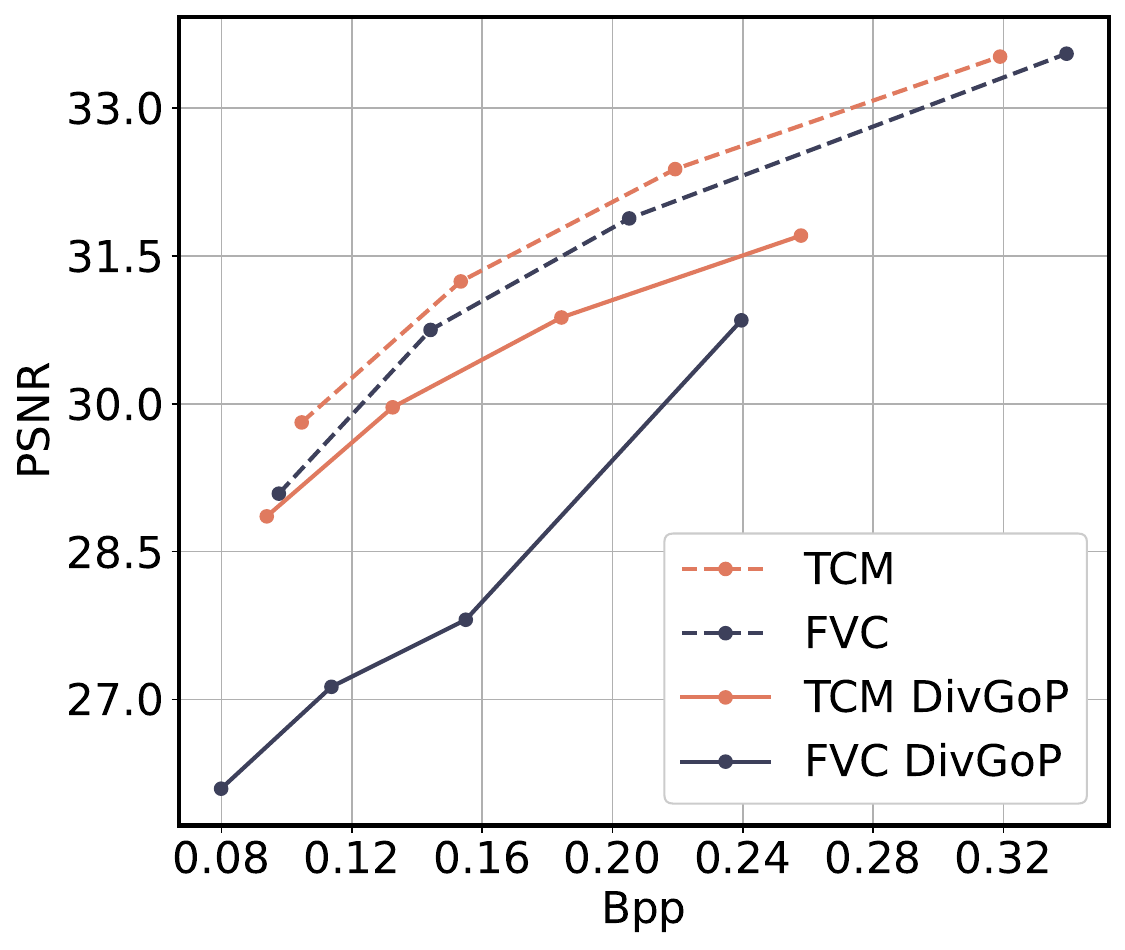}}
  \vspace{-0.1cm}
  \caption{Video Reconstruction Quality in terms of Bpp-PSNR on HEVC Class B dataset (Left) and Class C dataset (Right)}
  \label{fig:psnr}
\end{figure}

\begin{table}
\setlength{\tabcolsep}{4pt}
\footnotesize
  \caption{BD-RATE results on multi-object tracking.}
  \label{table:mot_bdbr}
  \centering
  \begin{tabular}{l|cccccc}
    \toprule 
   Method                       & MOTA   &  mAP & mAP50 & MOTP & FN & Average\\
   \hline
   DCVC \cite{li2021dcvc}       &0.0    & 0.0   &0.0    &0.0    & 0.0 & 0.0\\
   Ours+DCVC                    &-41.82 &-39.43 &-40.60 &-37.73 &-41.74 & \textbf{-40.26}\\
   \hline
   FVC \cite{hu2021fvc}         &-5.32  &-9.75  &-14.53 &-1.39  &-6.20 & -7.44\\
   Ours+FVC                     &-34.28 &-31.27 &-32.09 &-35.02 &-32.89 & \textbf{-33.11}\\
   \hline
   TCM \cite{sheng2022temporal} &-25.19 &-32.34 &-31.02 &-39.85 &-26.44 & -30.97\\
   Ours+TCM                     &-40.82 &-45.10 &-46.15 &-51.66 &-38.98 & \textbf{-44.54}\\
   \hline
   HEVC \cite{hevc}             &-33.88 &-31.35 &-40.43 &15.80  &-34.02 & -24.78\\
    \bottomrule
  \end{tabular}
\end{table}

\subsection{Video Action Recognition}
We conducted an evaluation of our method on the UCF101 dataset, benchmarking against TCM and FVC codecs. The outcomes of this comparison are depicted in Fig.~\ref{fig:mot}(f). For the task of video action recognition, rather than initiating a new set of network trainings, we employed pre-trained mode prediction networks previously utilized for the MOT task and use the ``DivGoP" structure for validation. The original videos were compressed using the various codecs, and the resulting decoded videos served as inputs to the TSM model as cited in Lin et al.~\cite{lin2019tsm}. The experimental results indicate that our method maintains efficacy in this context, securing approximately 21.81\% and 26.03\% in bitrate savings for TCM and FVC, respectively.


\subsection{Video Reconstruction Quality}

When it comes to the performance of reconstructing the video for human viewing, the situation is straightforward for conventional video codecs, no matter learned or traditional ones: just encode the frames, transmit and decode the compressed bitstream, and measure their quality using appropriate metrics (PSNR, MS-SSIM, etc). In our method, there're two options: Control the encoding procedure for human viewing, which is the ideal way for best frame reconstruction quality, or just continue the encoding procedure used for machine vision tasks. 

Note that our method does not change the weights of the original encoder and decoder in pre-trained DVC, so using this option can directly recover the human viewing quality to original best performance of the codecs. On the other hand, we also evaluate the human viewing quality of the encoded bitstream for machine, using ``DivGop" encoding structure on HEVC Class B and Class C test datasets. These are referred to as ``FVC DivGoP" and ``TCM DivGoP" in Fig.~\ref{fig:psnr}. Results show that our control for machine does decrease the video reconstruction quality of the original TCM codec by about 1dB in PSNR. Besides, it is also observed that the PSNR of residual codec FVC drops more sharply than that of conditional codecs like TCM. However, these are acceptable since we can always adjust the encoder to compress the video for human viewing when it's needed.

\subsection{Ablation Study}

\begin{figure}[t]
  \subfloat
  {\includegraphics[scale=0.22]{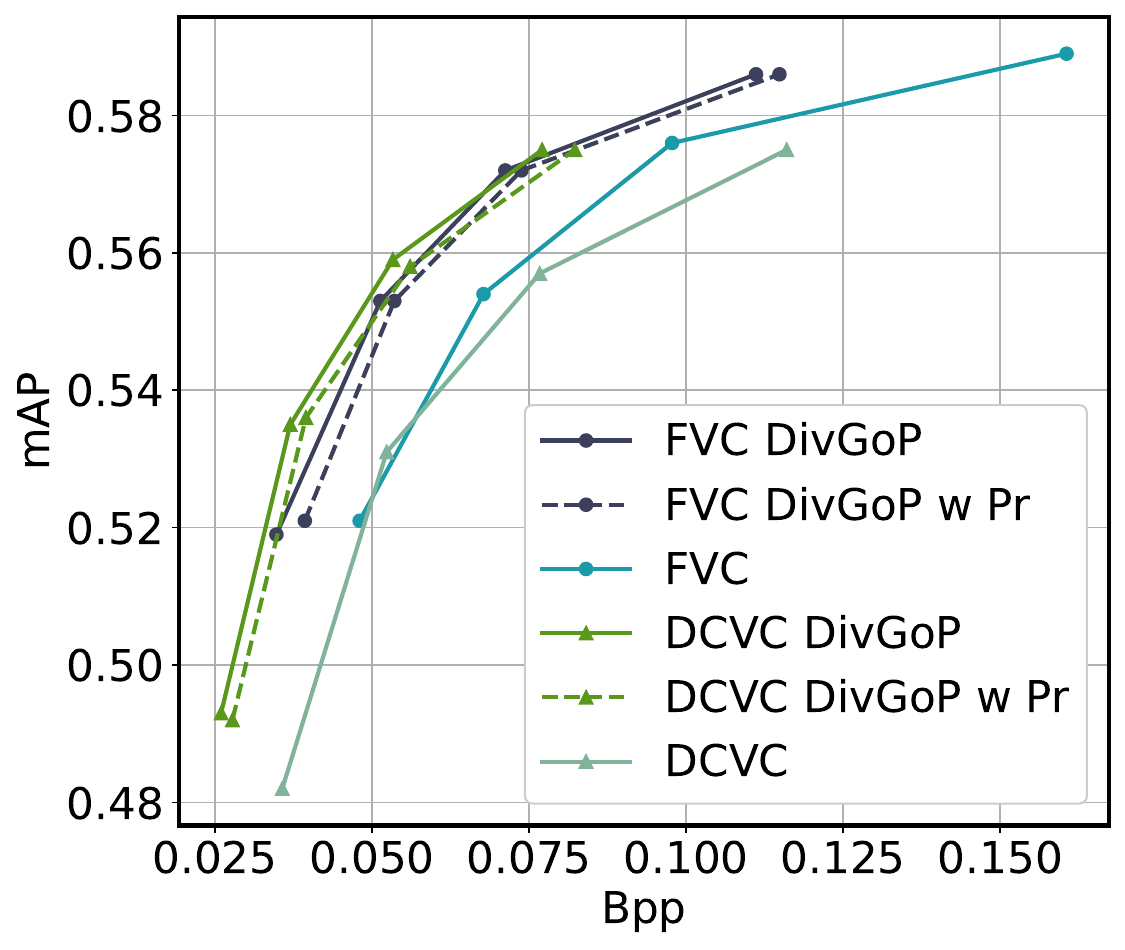}}
  \subfloat
  {\includegraphics[scale=0.22]{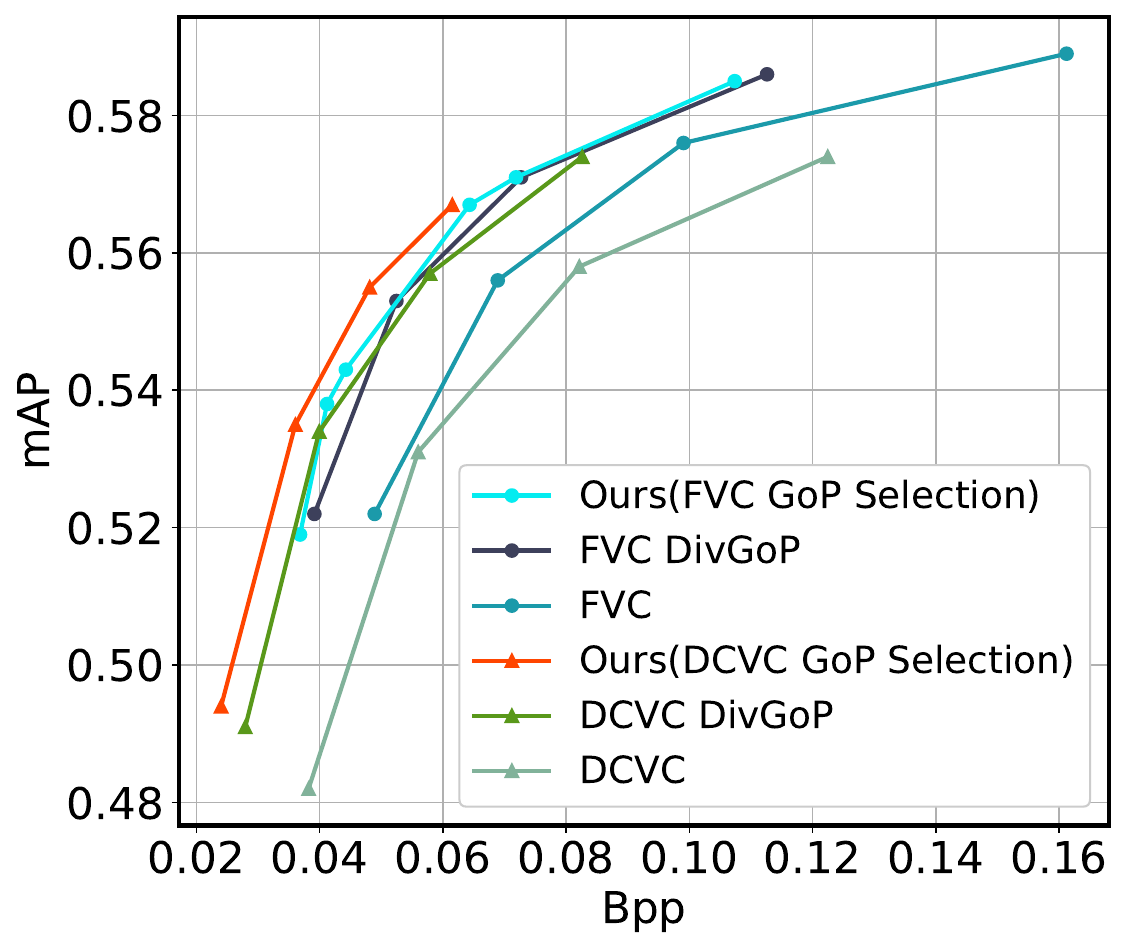}}
  \caption{Left: Ablation study for Mode Prediction network. Right: Ablation study for GoP Selection network}
  \label{fig:ablation}
  \vspace{-0.1cm}
\end{figure}

\begin{figure}[t]
  \centering
  \subfloat
  {\includegraphics[scale=0.4]{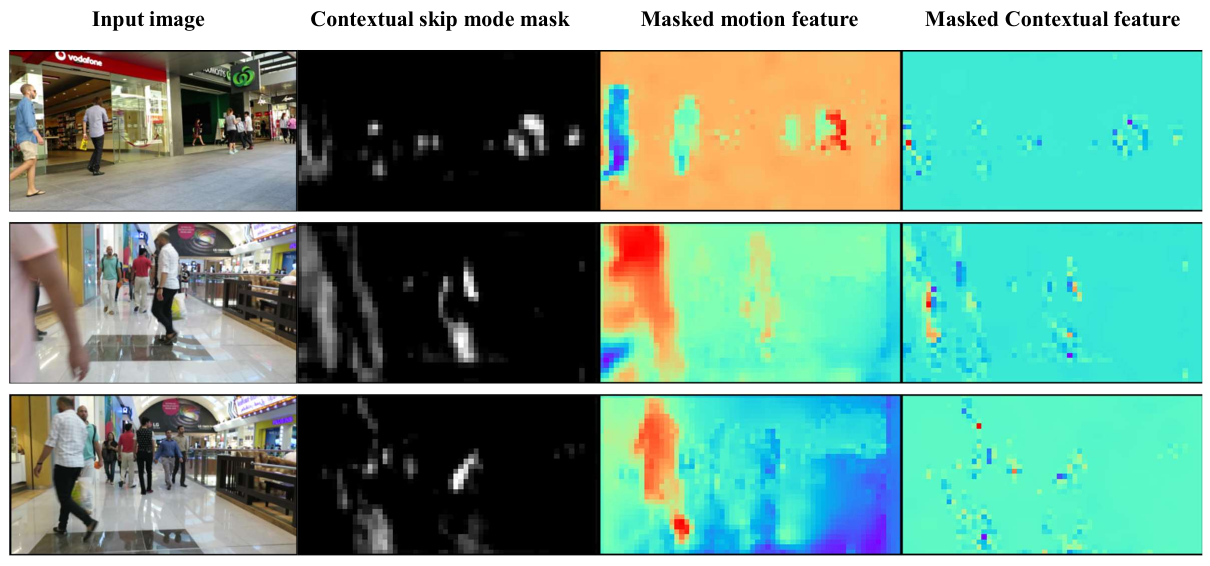}}
  \caption{Visualization results of the mode prediction networks.}
  \vspace{-0.4cm}
  \label{fig:visualization}
\end{figure}
\textbf{Mode Prediction}.
In our method, we propose mode prediction network to decide the coding modes of motion and residual/contextual feature elements for machine vision, mainly targeting on reducing the bitrate and keep the crucial information for important objects in the frame. Moreover, mainstream DVC, including FVC, DCVC, TCM and others, use motion estimation and warping operation to generate a rough predicted frame, which we call $P_r$ frame here. The $P_r$ frame also meets our requirement of “consuming lower code streams but retaining object and motion information” to a certain extent. So comparing it with the $P_m$ frame used in our method is also a point worthy of attention. In ``DivGoP" structure, we use the $P_r$ frame to replace the original $P_m$ frames, and the GoP structure will be $[I, P_r, P, P_r,...]$. As shown in the left figure in Fig.~\ref{fig:ablation}, based on DCVC and FVC, we compare the structure using $P_r$ frames with the original ``DivGoP" structure using $P_m$ frames, corresponding to the ``DCVC/FVC DivGoP w $P_r$" and ``DCVC/FVC DivGoP" curves respectively. It is observed that the ``DivGoP w $P_r$" curve also achieves about 28.68\% and 16.24\% bitrate savings for DCVC and FVC, respectively. But our ``DivGoP" curve shows better trade-off, by achieving 32.67\% and 19.43\% bitrate savings.

\noindent\textbf{GoP Selection}.
We prove the effectiveness of Gop Selection module in terms of rate-precision performance on MOT dataset. 
As shown in the right figure in Fig.~\ref{fig:ablation}, we compare the Bpp-mAP performance of ``DCVC/FVC Gop Selection" and ``DCVC/FVC DivGoP" structure on MOT Datast. Using DCVC as an example, experiment results show that our Gop Selection achieves about 39.43\% bitrate saving, which does show better trade-off than that 32.67\% bitrate saving of hand-crafted ``DivGoP" structure. 

\noindent\textbf{Simply fine-tuning DVC for machine vision tasks}.
We employed the loss function $L_f = R + \lambda_1 MSE + \lambda_2 L_{det}$ to fine-tune the FVC encoder and the entire FVC codec for the MOT task. As shown in Fig.~\ref{fig:dfs} right, while simple fine-tuning improves accuracy, it also notably increases the bitrate. This phenomenon could be attributed to the degradation in the fidelity of decoded frames when optimized for downstream tasks. Due to the inter-frame reference nature of video codecs, this degradation is propagated across frames, resulting in a high bitrate consumption. This observation aligns with findings reported in DeepSVC~\cite{lin2023deepsvc}, indicating that such task-specific performance can not be achieved with simply finetuning a DVC. This also reflects the superiority of our method.

\noindent\textbf{Visualization Results}.
We present the visualization results of our mode prediction network, as shown in Fig.~\ref{fig:visualization}. The first column shows the input images, the second column shows the skip mode masks for contextual features (averaged by channel), and the third and fourth columns are masked motion and contextual features, respectively, where the red color indicates large value and green color indicates small value. It is observed that the predicted coding mode mask for contextual features kept the regions of the small objects in the frames, and masked the conspicuous objects which are easy to recognize and the background regions. This shows the network's preference for objects in the frame. As for the motion information, it is observed that the predicted masks kept most of the elments of moving objects, which are crucial for detection and tracking.

\noindent\textbf{Complexity and Encoding Latency Analysis}.
The parameter numbers of our dynamic vision mode prediction network and GoP selection network are 1.48M and 5.36M. 
Since our method controls the video encoder during the encoding phase, such an operation brings encoding latency. To measure the impact on the encoding phase, we include a comparison of average encoding latency versus performance on MOT Dataset ($960\times512$) in Tab.~\ref{tab:latency}. While our method bring an increase in encoding latency (1.738s vs. 1.675s), the improvement in R-P performance is substantial. Besides, the decoding procedure and decoder side models are not changed, which brings convenience to the deployment of the decoder and support for multiple tasks.

\begin{table}
    \centering
    \setlength{\tabcolsep}{2pt}
    \vspace{-7mm}
    \caption{Encoding Latency vs. Performance.}
    \label{tab:latency}
    \begin{tabular}{l|cc}
        \toprule 
        Method & Encoding Latency(s) & BD-BR Performance \\ \hline
        
        DCVC & 1.675 & 0.0 \\ 
        Ours+DCVC & 1.738 & \textbf{-40.26} \\
        \hline
        FVC & 0.116 & -7.44 \\ 
        Ours+FVC & 0.159 & \textbf{-33.11} \\ 
        \bottomrule
    \end{tabular}
\vspace{-5mm}
\end{table}

\section{Conclusion}

In this paper, we propose a flexible framework for deep video compression models, where the pre-trained encoders can be controlled to change the encoding pipeline for machine vision tasks, achieving significant better bpp-precision trade-off without changing the decoders or decoding procedure. The controlling methods are attributed to two main developments, including dynamic vision mode prediction network and GoP structure selection network. Experiment results show that our framework is general for residual and conditional deep video codecs and different downsteam vision tasks like detection, tracking and action recognition. Our comprehensive evaluation, conducted against existing deep video codecs, confirms the superior performance of our models, establishing a new benchmark for future deep video coding for machine. The effectiveness of each component is clearly evidenced through extensive ablation studies. 

\noindent \textbf{Acknowledgments.} 
This work was supported by the National Natural Science Fund of China (Project No. 42201461) and the General Research Fund (Project No. 16209622) from the Hong Kong Research Grants Council.

{
    \small
    \bibliographystyle{ieeenat_fullname}
    \bibliography{main}
}

\end{document}